\documentclass[10pt]{revtex4}
\usepackage{graphicx}

\newcount\driver
\newcount\bozza

\font\ottorm=cmr8\font\ottoi=cmmi8\font\ottosy=cmsy8%
\font\ottocss=cmcsc8%
\font\sixrm=cmr6\font\sixi=cmmi6\font\sixsy=cmsy6%
\font\fiverm=cmr5\font\fivesy=cmsy5
\font\fivei=cmmi5
\font\tenmib=cmmib10
\font\sevenmib=cmmib10 scaled 800

 2

\font\sc=cmcsc10

\font\elevenrm=cmr11
\font\twelverm=cmr12
\font\ottorm=cmr8
\font\msytw=msbm9 scaled\magstep1

\font\indbf=cmbx10 scaled\magstep2

\font\ottorm=cmr8\font\ottoi=cmmi8\font\ottosy=cmsy8%
\font\ottocss=cmcsc8%
\font\sixrm=cmr6\font\sixi=cmmi6\font\sixsy=cmsy6%
\font\fiverm=cmr5\font\fivesy=cmsy5
\font\fivei=cmmi5

\def\ottopunti{\def\rm{\fam0\ottorm}%
\textfont0=\ottorm\scriptfont0=\sixrm\scriptscriptfont0=\fiverm%
\textfont1=\ottoi\scriptfont1=\sixi\scriptscriptfont1=\fivei%
\textfont2=\ottosy\scriptfont2=\sixsy\scriptscriptfont2=\fivesy%
\textfont4=\ottocss\scriptfont4=\sc\scriptscriptfont4=\sc%
\scriptfont4=\ottocss\scriptscriptfont4=\ottocss%
\textfont5=\tenmib\scriptfont5=\sevenmib\scriptscriptfont5=\fivei
\setbox\strutbox=\hbox{\vrule height7pt depth2pt width0pt}%
\normalbaselineskip=9pt\let\sc=\sixrm\normalbaselines\rm}

\mathchardef\BDpr = "0540  
\mathchardef\Bg   = "050D  

{\count255=\time\divide\count255 by 60 \xdef\hourmin{\number\count255}
        \multiply\count255 by-60\advance\count255 by\time
   \xdef\hourmin{\hourmin:\ifnum\count255<10 0\fi\the\count255}}

\def\openone{\leavevmode\hbox{\elevenrm 1\kern-3.63pt\twelverm1}}%
\def\*{\vglue0.5truecm}

 \let\b=\beta     
\let\z=\zeta       \let\l=\lambda
\let\m=\mu             \let\p=\pi    
\let\s=\sigma \let\t=\tau

\def\\{\hfill\break} \let\==\equiv

\let\io=\infty 

\let\0=\noindent

\def\tende#1{\,\vtop{\ialign{##\crcr\rightarrowfill\crcr
 \noalign{\kern-1pt\nointerlineskip} \hskip3.pt${\scriptstyle
 #1}$\hskip3.pt\crcr}}\,}
\def\circage{\lower2pt\hbox{$\,\buildrel > \over {\scriptstyle \sim}\,$}}
\def\otto{\,{\kern-1.truept\leftarrow\kern-5.truept\to\kern-1.truept}\,}

 \def\VV{{\cal V}}

\def\T#1{{#1_{\kern-3pt\lower7pt\hbox{$\widetilde{}$}}\kern3pt}}
\def\VVV#1{{\VV #1}_{\kern-3pt
\lower7pt\hbox{$\widetilde{}$}}\kern3pt\,}
\def\W#1{#1_{\kern-3pt\lower7.5pt\hbox{$\widetilde{}$}}\kern2pt\,}

\def\indica{\leaders \hbox to 0.5cm{\hss.\hss}\hfill}
\def\guida{\leaders\hbox to 1em{\hss.\hss}\hfill}

\def\hhh{{\bf h}}

\def\ul{\underline}

\mathchardef\dd   = "050E
\mathchardef\aa   = "050B
\mathchardef\bb   = "050C
\mathchardef\ggg  = "050D
\mathchardef\xxx  = "0518
\mathchardef\zzzzz= "0510
\mathchardef\oo   = "0521
\mathchardef\lll  = "0515
\mathchardef\mm   = "0516
\mathchardef\Dp   = "0540
\mathchardef\H    = "0548
\mathchardef\FFF  = "0546
\mathchardef\ppp  = "0570
\mathchardef\Bn   = "0517
\mathchardef\pps  = "0520
\mathchardef\fff  = "0527
\mathchardef\FFF  = "0508
\mathchardef\nnnnn= "056E

\def\to{\rightarrow}

\def\qed{\hfill\raise1pt\hbox{\vrule height5pt width5pt depth0pt}}

\def\indic{\hbox{\raise-2pt \hbox{\indbf 1}}}

\def\TTT{\hbox{\msytw T}}

\def\ul#1{{\underline#1}}

\def\V0{{\bf 0}}

\font\tenmib=cmmib10 
\font\sevenmib=cmmib7\font\fivemib=cmmib5

\font\fivei=cmmi5\font\sixi=cmmi6\font\ottoi=cmmi8
\font\ottorm=cmr8\font\fiverm=cmr5\font\sixrm=cmr6
\font\ottosy=cmsy8\font\sixsy=cmsy6\font\fivesy=cmsy5
\font\ottocss=cmcsc8%

\mathchardef\Ba   = "050B  
\mathchardef\Bb   = "050C  
\mathchardef\Bg   = "050D  
\mathchardef\Bd   = "050E  
\mathchardef\Be   = "0522  
\mathchardef\Bee  = "050F  
\mathchardef\Bz   = "0510  
\mathchardef\Bh   = "0511  
\mathchardef\Bthh = "0512  
\mathchardef\Bth  = "0523  
\mathchardef\Bi   = "0513  
\mathchardef\Bk   = "0514  
\mathchardef\Bl   = "0515  
\mathchardef\Bm   = "0516  
\mathchardef\Bn   = "0517  
\mathchardef\Bx   = "0518  
\mathchardef\Bom  = "0530  
\mathchardef\Bp   = "0519  
\mathchardef\Br   = "0525  
\mathchardef\Bro  = "051A  
\mathchardef\Bs   = "051B  
\mathchardef\Bsi  = "0526  
\mathchardef\Bt   = "051C  
\mathchardef\Bu   = "051D  
\mathchardef\Bf   = "0527  
\mathchardef\Bff  = "051E  
\mathchardef\Bch  = "051F  
\mathchardef\Bps  = "0520  
\mathchardef\Bo   = "0521  
\mathchardef\Bome = "0524  
\mathchardef\BG   = "0500  
\mathchardef\BD   = "0501  
\mathchardef\BTh  = "0502  
\mathchardef\BL   = "0503  
\mathchardef\BX   = "0504  
\mathchardef\BP   = "0505  
\mathchardef\BS   = "0506  
\mathchardef\BU   = "0507  
\mathchardef\BF   = "0508  
\mathchardef\BPs  = "0509  
\mathchardef\BO   = "050A  
\mathchardef\BDpr = "0540  
\mathchardef\Bstl = "053F  

\let\aa=\Ba\let\fff=\Bf
\let\oo=\Bo\let\nn=\Bn
\let\pps=\Bps\def\hhh={\V h}
\let\bb=\Bb

\def\TTT{\hbox{\msytw T}}

\let\ul=\underline

\def\ins#1#2#3{\vbox to0pt{\kern-#2 \hbox{\kern#1 #3}\vss}\nointerlineskip}
\newdimen\xshift \newdimen\xwidth \newdimen\yshift
\newcount\griglia

\def\insertplot#1#2#3#4#5#6{%
\xwidth=#1pt \xshift=\hsize \advance\xshift by-\xwidth \divide\xshift by 2%
\begin{figure}[ht]
\vspace{#2pt}
\hspace{\xshift}
\begin{minipage}{#1pt}
#3
\ifnum\driver=1 \griglia=#6
\ifnum\griglia=1
\openout13=griglia.ps
\write13{gsave .2 setlinewidth}
\write13{0 10 #1 {dup 0 moveto #2 lineto } for}
\write13{0 10 #2 {dup 0 exch moveto #1 exch lineto } for}
\write13{stroke}
\write13{.5 setlinewidth}
\write13{0 50 #1 {dup 0 moveto #2 lineto } for}
\write13{0 50 #2 {dup 0 exch moveto #1 exch lineto } for}
\write13{stroke grestore}
\closeout13
\includegraphics{griglia.ps}
\fi
\includegraphics{#4}\fi%
\end{minipage}
\caption{#5}
\end{figure}
}

\def\lb#1{%
\ifnum\bozza=1
\label{#1}\rlap{\kern6.truecm{$\scriptstyle#1$}}
\else\label{#1}
\fi}

\def\be{\begin{equation}}
\def\ee{\end{equation}}
\def\bea{\begin{eqnarray}}\def\eea{\end{eqnarray}}
\def\bean{\begin{eqnarray*}}\def\eean{\end{eqnarray*}}
\def\bfr{\begin{flushright}}\def\efr{\end{flushright}}
\def\bc{\begin{center}}\def\ec{\end{center}}
\def\ba#1{\begin{array}{#1}} \def\ea{\end{array}}
\def\bd{\begin{description}}\def\ed{\end{description}}
\def\bv{\begin{verbatim}}\def\ev{\end{verbatim}}
\def\nn{\nonumber}
\def\Halmos{\hfill\vrule height10pt width4pt depth2pt \par\hbox to \hsize{}}

\newdimen\xshift \newdimen\xwidth \newdimen\yshift \newdimen\ywidth

\def\ins#1#2#3{\vbox to0pt{\kern-#2\hbox{\kern#1 #3}\vss}\nointerlineskip}

\def\eqfig#1#2#3#4#5{
\par\xwidth=#1 \xshift=\hsize \advance\xshift
by-\xwidth \divide\xshift by 2
\yshift=#2 \divide\yshift by 2
\line{\hglue\xshift \vbox to #2{\vfil
#3 \includegraphics{#4.ps}
}\hfill\raise\yshift\hbox{#5}}}

\def\8{\write12}

\begin{document}

\title{Fluctuation Theorem, non linear response and the regularity of time reversal symmetry}
\*
\author{Marcello Porta}
\affiliation{Dipartimento di Fisica,
Universit\`a di Roma ``La Sapienza'', Piazzale Aldo Moro 2, 00185 Roma Italy}

\begin{abstract}
The Gallavotti -- Cohen Fluctuation Theorem (FT) implies an infinite
set of identities between correlation functions that can be seen as a
generalization of Green Kubo formula to the nonlinear regime. As an application, we discuss a perturbative check of the FT relation through these identities for a simple Anosov reversible system; we find that the lack of differentiability of the time reversal symmetry implies a {\em violation} of the Gallavotti -- Cohen fluctuation relation. Finally, a brief comparison with Lebowitz --  Spohn FT is reported.
\end{abstract}

\maketitle

\section{Introduction}\label{sec1}

Despite many proposals have been advanced, a general theory of the steady state of dissipative systems is still lacking. Nevertheless, it is a remarkable fact that {\em under suitable hypothesis} something can be said, and physical predictions can be made; for example, chaotic hypothesis (CH), \cite{Gsch}, stating that {\em for the purpose of studying macroscopic properties, a system exhibiting chaotic motions may be regarded as a transitive hyperbolic (that is, {\em Anosov}) one}, see \cite{ergo}, implies two remarkable results: the Gallavotti -- Cohen fluctuation theorem, \cite{GC95a}, \cite{Gfluttsch}, holding for {\em reversible} systems, and a formula describing the linear response of nonequilibrium systems, due to Ruelle, \cite{R2lin}, \cite{Rdiff}. These results have been and are still widely studied in the physical literature; see \cite{ECM93} (where a relation which inspired the FT was empirically discovered), \cite{Zflutt}, \cite{BCG}, \cite{toychaos}, \cite{henon}, \cite{Tou} for instance. In this note we focus on the first on these two results; in particular, in section \ref{sec2} we show that FT has some nice implication on nonlinear response (in \cite{G96Ons} it has been already pointed out that FT implies the usual linear response theory, that is Green Kubo formula (GK) and Onsager reciprocity relations), while in section \ref{sec3} we discuss a check of FT in a simple Anosov reversible system. Interestingly, a simple perturbative calculation shows that the lack of differentiability of the time reversal symmetry operator implies a {\em violation} of the Gallavotti -- Cohen fluctuation relation. Finally, in section \ref{sec4} we show that in presence of a non differentiable time reversal symmetry an identity equivalent to the result proved by Lebowitz and Spohn in \cite{LS} is true.
 
The connection between FT and nonlinear response consists in the fact that FT implies identities between correlation functions of physical observables in a nonequilibrium steady state; this has been pointed out first in \cite{AG}, where the authors considered systems whose evolution was stochastic and ruled by a master equation, and independently in \cite{tesi}, in the context of deterministic systems satisfying CH. The results of \cite{tesi} are presented in sections \ref{sec2} and \ref{sec3} of this paper.

Before turning to our results, we spend a few words on some of the main features of Anosov systems; we refer the interested reader to \cite{ergo} for a modern introduction to the subject. Consider a generic discrete dynamical system $x_{k} = S^{k} x_{0}$ (discreteness only causes technical problems, see \cite{Ge96}, since $S$ can be thought as the map arising from the Poincar\' e section of a system evolving in continuous time), and assume that $S$ is Anosov; informally, this means that given a point $x$ the nearby points separate exponentially fast from $x$ in the future and in the past, except when located on a surface $W_{s}(x)$ ({\em stable} manifold) or $W_{u}(x)$ ({\em unstable} manifold), respectively for the future and for the past.

It is a well known result that Anosov systems admit an invariant measure $\m_+$, the so -- called {\em SRB} measure; in fact, given a sufficiently regular observable $F(x)$ the following equality holds:
\be
\lim_{T\rightarrow+\infty}\frac{1}{T}\sum_{j=0}^{T-1}F(S^{j}x) = \int \mu_{+}(dx)F(x)\;,\label{1.1}
\ee
{\em apart} from a set of points of zero volume measure. It is a remarkable fact that the SRB measure admit in principle an {\em explicit} representation, similar to the equilibrium Gibbs distribution.

 Notice that at equilibrium, that is when the system is stationary and non dissipative, the chaotic hypothiesis implies the ergodic hypothesis, in the sense that assuming CH the SRB measure is the Liouville one; but in general, when dissipation in present the SRB measure is {\em singular} with respect to the volume, that is it is concentrated on a zero volume set.

 To conclude, the SRB distribution verifies a large deviation theorem (see for example \cite{Ge96} for a proof of this statement for a special choice of $F$ and in the more complex case of Anosov flows). In fact, consider the finite time average $f = \frac{1}{\t}\sum_{j=-\frac{\t}{2}}^{\frac{\t}{2} - 1}F(S^{j}x)$; then, it is possible to prove that there are values $f_1,f_2$ such that if $[a,b]\in(f_1,f_2)$ then $Prob_{\m_+}(f\in[a,b])\sim e^{\t\z_{F}(f)}$, in the sense that
\be
\lim_{\t\rightarrow+\infty}\frac{1}{\t}\log Prob_{\m_{+}}(f\in[a,b]) = \max_{f\in[a,b]}\z_{F}(f)\;,\lb{1.2}
\ee
 and $\z_{F}(f)$ is analytic and convex in $(f_{1},f_{2})$.

\section{Fluctuation theorem and nonlinear response}\label{sec2}

At present time, no universally accepted definition of entropy for a dissipative system has been given. Nevertheless, the {\em rate of entropy production} is a well defined quantity, and is proportional to the work per unit time made by the thermostats on the system; the proportionality factor is the inverse of the temperature of the termostats (setting to $1$ the Boltzmann constant). In particular, for a special class of thermostats, the {\em gaussian} ones, the entropy production rate corresponds to the {\em phase space contraction}, that is to minus the divergence of the equation of motion, see \cite{Gordch}; this fact can be taken as a general definiton of entropy production, if one assumes that the steady state of a large system is not affected by the details of the termostatting mechanism which ensures the existence of the steady state.

In the case of a system evolving in discrete time, which is the case that we want to consider, the entropy production rate $\s$ is given by $\s(x) = -\log|\det \partial S(x)|$, where $\partial S$ if the jacobian of the time evolution $S$.

Now, let $p$ be the adimensional average over a time $\t$ of $\s(x)$, that is $p = \frac{1}{\t\sigma_{+}}\sum_{j=-\frac{\t}{2}}^{\frac{\t}{2}-1}\s(S^{j}x)$, where $\s_+$ is the SRB expectation of $\s(x)$, {\em i.e.} $\s_{+} = \int \m_{+}(dx)\s(x)$, and call $\z(p)$ the large deviation functional of $\s$, as defined in (\ref{1.2}); assume that CH holds, and that the system is {\em reversible}, which means that there exists a differentiable isometry $I$ such that $I\circ S = S^{-1}\circ I$, $I\circ I = 1$. Then, as proven by Gallavotti and Cohen, see \cite{GC95a} or \cite{GdimFT} for a proof detailed from a formal viewpoint, the following result holds.
\vskip.2cm
{\bf Fluctuation theorem (FT):} {\em There is $p^{*}\geq 1$ such that for $|p|< p^{*}$}
\be
\z(-p) = \z(p) -p\s_{+}\;,\lb{2.1}
\ee
\vskip.2cm

This result has an interesting corollary. Setting $\pi_{\t}(q)dq = Prob_{\m_+}(p\in[q,q+dq])$, define $\l(\b)$ as
\be
\l(\b) = \lim_{\t\rightarrow +\infty}\frac{1}{\tau}\log\int e^{\t(q-1)\s_{+}\beta} \pi_{\t}(q)dq\;;\lb{2.2}
\ee
clearly, $\l(\b)$ is related to $\z(p)$ through a Legendre transform, that is
\be
\z(p) = \max_{\b}\left(\b\s_+(p-1) - \l(\b) \right)\;,\lb{2.3}
\ee
and moreover $\l(\b)$ admits the following expansion:
\be
\l(\b) = \sum_{n\geq 2}\sum_{t_1,...,t_{n-1} = -\infty}^{+\infty}<\s(\cdot)\s(S^{t_1}\cdot)\,...\,\s(S^{t_{n-1}}\cdot)>^{T}_{+}\frac{\beta^{n}}{n!} \equiv \sum_{n\geq 2}C_{n}\frac{\b^{n}}{n!}\;,\lb{2.4}
\ee
where by $<...>^{T}_{+}$ we denote the {\em cumulant} with respect to the SRB measure $\m_{+}$. It is straightforward to see that the fluctuation theorem implies an identity for the generating functional $\l(\b)$, see \cite{LS}, \cite{AG}, \cite{Tou} and \cite{tesi} for instance: it follows that, as a consequence of the relation $\p_{\t}(p) \sim e^{\t p\s_+}\pi_{\t}(-p)$, valid under the hypothesis of FT, 
\be
\l(\b) = \l(-1-\b) - \s_{+}(2\b + 1)\;.\lb{2.5}
\ee
Notice that the generating functional of the cumulants is usually defined as (see \cite{LS}, \cite{AG}, \cite{Tou}) $\tilde \l(\b) = \lim_{\t\rightarrow+\infty}\frac{-1}{\tau}\log\left< e^{-\t p\s_+ \b } \right>$, and with this definition the relation (\ref{2.5}) is replaced by the more familiar $\tilde \l(\b) = \tilde \l(1 - \b)$; but the two definitions are equivalent, since $\l(\b) = -\b\s_+ - \tilde\l(-\b)$. Formula (\ref{2.5}) translates immediatly in a relation for $\s_+$; in fact, (\ref{2.5}) evaluated at $\b=0$ becomes:
\be
0 = \l(-1) -\s_{+}\Rightarrow \s_{+} = \sum_{n\geq 2}C_{n}\frac{(-1)^{n}}{n!}\;.\lb{2.6}
\ee
Assuming that the entropy production $\s(x)$ has the form $\s(x) = \sum_i G_{i}J^{(0)}_{i}(x) + O(G^{2})$, where $\{G_{i}\}$, and $\{J^{(0)}_{i}(x)\}$ are respectively the forcing parameters and the corresponding currents, it has been shown in \cite{G96Ons} that the identity
\be
\sum_{i,j}G_{i}G_{j}\partial_{G_{i}G_{j}}\s_{+}\big|_{\ul G=\ul 0} = \sum_{i,j}G_{i}G_{j}\partial_{G_{i}G_{j}}C_{2}\big|_{\ul G=\ul 0}\;,\lb{2.7}
\ee
which is nothing else that (\ref{2.6}) at second order, is equivalent to the Green Kubo formula, stating that
\be
L_{ij}\equiv \partial_{G_{j}}<J^{(0)}_{i}>_{+}\big|_{\ul G=\ul 0} = \frac{1}{2}\sum_{t=-\infty}^{+\infty}<J^{(0)}_{i}(S^{t}\cdot)J^{(0)}_{j}(\cdot)>_{0}\;,\lb{2.8}
\ee
where by $<\cdot >_{0}$ we denote the expectation with respect to the invariant measure at zero forcing (which by CH is the Liouville measure). Hence, formula (\ref{2.6}) can be seen as a {\em generalization of GK formula to the nonlinear regime}, being an identity for $\s_+$ valid for $\ul{G}\neq \ul{0}$. Moreover, by taking derivatives with respect to $\b$ in the {\em r.h.s.} and in the {\em l.h.s.} of (\ref{2.5}) we find that
\be
C_{n} = \sum_{k\geq 0}\frac{(-1)^{k+n}C_{k+n}}{k!}\;\qquad n\geq 2\;,\lb{2.9}
\ee
which is a nontrivial identity for the cumulants valid at $\ul{G}\neq \ul{0}$.
Finally, these identities can be considerably extended by using a generalized version of FT. Consider a generic observable $O$ odd under time reversal symmetry, {\em i.e.} such that $O(Ix) = -O(x)$, let $w$ be its adimensional average over a finite time $\t$, that is $w= \frac{1}{\t O_+}\sum_{j=-\frac{\t}{2}}^{\frac{\t}{2} -1}O(S^{j}x)$ where $O_+$ is the SRB average of $O$, and call $\z(p,w)$ the large deviation functional of the joint probability distribution $\pi_\t(p,w)$ of $p$ and $w$; then, under the same hypothesis of FT the following result holds, as a special case of a much more general result in \cite{fluttrev}.
\vskip.2cm
{\bf Generalized Fluctuation Theorem (GFT):} {\em There are $w^*\geq 1$, $p^*\geq 1$ such that for $|p|< p^*$ and $|w|< w^*$
\be
\z(-p,-w) = \zeta(p,w) - p\s_+\;.\lb{2.10}
\ee
}
\vskip.2cm
One can define the generating functional $\l(\b_1,\b_2)$ of the mixed cumulants of $O$, $\s$ in a way analogous to (\ref{2.2}),
\be
\l(\b_1,\b_2) = \lim_{\t\rightarrow +\infty}\frac{1}{\tau}\log\int e^{\t(q-1)\s_{+}\beta_1 + \t(t - 1)O_+\b_2} \pi_{\t}(q,t)dq dt\;;\lb{2.11}
\ee
again, $\l(\b_1,\b_2)$ is related to $\z(p,w)$ through a Legendre transform, and moreover it can be expressed as
\be
\l(\b_1,\b_2) = \sum_{k\geq 2}\sum_{m,n\geq 0:\, m+n = k}\frac{\b_1^n \b_2^m}{n!m!}C_{n,m}\,\lb{2.12}
\ee
where $C_{n,m}=\partial_{\b_1}^{n}\partial_{\b_2}^{m}\l(\b_1,\b_2)\big |_{\ul\b=\ul 0}$, that is $C_{n,m}$ is given by a sum over times of mixed cumulants of $\s(S^{t_i}x),O(S^{t_j}x)$, $0\leq i\leq n$, $0\leq j\leq m$. Then, in full analogy with what has already been discussed, it is easy to show that GFT translates into an identity for the generating functional, namely
\be
\l(\b_1,\b_2) = \l(-1-\b_1,-\b_2) - (2\b_1 + 1)\s_+  - 2\b_2O_+\;,\lb{2.13}
\ee
which implies the following relations:
\bea
O_{+} &=& \sum_{k\geq 2}\frac{(-1)^{k}}{2^{k-1}}\sum_{l=0,\,2l\leq k-1}\frac{C_{k-(2l+1),2l+1}}{(2l+1)!(k-(2l+1))!}\;\lb{2.14}\\
C_{l,n-l}&=&\sum_{k\geq 0}\frac{(-1)^{n+k}}{k!}C_{k+l,n-l},\qquad n\geq 2\;,\;l\leq n\;; \lb{2.15}
\eea
equation (\ref{2.14}) is obtained setting $\b_1=\b_2=-\frac{1}{2}$ in (\ref{2.13}), while (\ref{2.15}) can be proved differentiating with respect to $\b_1$, $\b_2$ the {\em r.h.s.} and the {\em l.h.s.} of (\ref{2.13}). It is interesting to see what happens to formula (\ref{2.14}) in the linear regime. Assuming that $\s(x) = \sum_i G_{i}J^{(0)}_{i}(x) + O(G^{2})$ we can rewrite (\ref{2.14}) as
\bea
O_{+} &=& \sum_{k\geq 2,\,k\,even}\frac{1}{(k-1)!}\frac{(-1)^k}{2^{k-1}}C_{1,k-1} + \sum_{k\geq 3,\,k\,odd}\frac{1}{k!}\frac{(-1)^k}{2^{k-1}}C_{0,k} + O(G^2)\nn\\
&=& \frac{1}{2}C_{1,1} + \sum_{k\geq 4,\,k\,even}\frac{1}{(k-1)!}\frac{(-1)^{k}}{2^{k-1}}C_{1,k-1} + \sum_{k\geq 3,\,k\,odd}\frac{1}{k!}\frac{(-1)^k}{2^{k-1}}C_{0,k}\nn\\
&=& \frac{1}{2}C_{1,1}+\sum_{k\geq 3,\,k\,odd}\frac{1}{k!}\frac{(-1)^{k}}{2^{k-1}}\left(C_{0,k} - \frac{1}{2}C_{1,k}\right) + O(G^2)\lb{2.16}\;,
\eea
and from (\ref{2.15}) we find that for $k$ odd $C_{0,k} - \frac{1}{2}C_{1,k} = O(G^2)$. Hence,
\be
O_{+} = \frac{1}{2}C_{1,1} + O(G^2) = \frac{1}{2}\sum_{t = -\io}^{\io}<\s(S^j x)O(x)>_+ + O(G^2)\;,\lb{2.17}
\ee
which gives:
\be
\partial_{G_i}O_+\big|_{\ul G=\ul 0} = \frac{1}{2}\sum_{t=-\io}^{\io}<J^{(0)}_i(S^t x)O(x)>_0\;.\lb{2.18}
\ee
Formula (\ref{2.18}) describes the linear response of a generic observable {\em odd} under time reversal symmetry; this result can be seen as a special case of the remarkable linear response formula obtained by Ruelle, \cite{R2lin}, valid in a much more general context.

\section{Check of Fluctuation Theorem for a simple Anosov system}\lb{sec3}
In this section we will perform a check of the fluctuation relation (\ref{2.1}) for a simple Anosov model, the  {\em perturbed Arnold cat map}, starting from the identites (\ref{2.6}), (\ref{2.9}); as we are going to see, the lack of differentiability of the time reversal simmetry operator implies a violation of (\ref{2.1}). Consider the following discrete evolution on the bidimensional torus $\TTT^{2}$
\be
\ul{x}_{k} = S^{k}_{\varepsilon}\ul{x}_{0}\qquad \mbox{mod $2\pi$}\;,\lb{3.1}
\ee
where
\be
S_{\varepsilon}\ul{x} = S\ul{x} + \varepsilon \ul{f}(\ul{x}),\qquad S \equiv \pmatrix{1 & 1 \cr 1 & 2},\quad \ul{f}(\ul{x}) = \pmatrix{f_{1}(\ul{x}) \cr f_{2}(\ul{x})}\;,\lb{3.2}
\ee
and $f_{1}(\ul{x})$, $f_{2}(\ul{x})$ are trigonometric polynomials. The map $S$ is the so --  called Arnod cat map, which is the simplest example of Anosov map: in fact, the eigenvalues $\l_+$, $\l_-$ of $S$ are such that $\l_+ > 1$, $\l_-<1$. Moreover the map $S$ is reversible, that is there exist $I$ such that $I\circ S = S^{-1}\circ I$, $I\circ I = 1$, where
\be
I = \pmatrix{-1& 0 \cr -1 & 1}.\lb{3.3}
\ee
Notice that since $S$ is conservative ($\det S = 1$) $\s_+ = 0$, which makes meaningless the fluctuation relation (\ref{2.1}) (the adimensional quantity $p$ is not defined); but one can derive the analogous of (\ref{2.1}) for the {\em dimensional} quantity $p'=\s_+ p$, and this relation becomes trivial if the evolution is conservative because in this case $p'=0$.

Consider now $\varepsilon\neq 0$. By the structural stability of Anosov systems, the evolution generated by $S_{\varepsilon}$ is still Anosov {\em provided} $\varepsilon$ is chosen small enough. In fact, for $\varepsilon<\varepsilon_0$ a {\em conjugation} $H_\varepsilon$, defined by the identity $S_{\varepsilon}\circ H_\varepsilon = H_\varepsilon \circ S$, can be explicitly constructed through a convergent power series in $\varepsilon$, and it turns out that $H_{\varepsilon}(\underline{x})$ is H\"older continuos in $\ul{x}$, \cite{ergo}. In an analogous way, the SRB measure can be explicitly constructed, and it follows that the expectations of H\"older continuous functions exists and are analytic in $\varepsilon$, \cite{ergo}. In particular it follows that, generically,

\be
\s_+  \equiv <-\log|\partial S_{\varepsilon}|>_+>0\;,\lb{3.4}
\ee
which means that the system is dissipative and so that the invariant measure is singular with respect to the volume; hence, the check of the fluctuation relation (\ref{2.1}) is nontrivial in this case. Notice that the proof of FT, \cite{GdimFT}, requires that the evolution is reversible, and in particular that the time reversal symmetry operator is differentiable; in our specific case the existence of $H_{\varepsilon}$ implies that $I_{\varepsilon} = H_{\varepsilon}\circ I \circ H_{\varepsilon}^{-1}$ verifies $I_{\varepsilon}\circ S_{\varepsilon} = S_{\varepsilon}^{-1}\circ I_\varepsilon$, {\em but} due to the mild regularity properties of $H_{\varepsilon}$, $I_{\varepsilon}(\ul x)$ is likely to be not differentiable. Then one can ask what happens to the fluctuation relation; to understand this, we make the choice $f_{1}(\ul{x}) = \sin(x_1) + \sin(2x_{1})$, $f_{2}(\ul{x}) =0$. An explicit computation shows that the linear response is still valid, as expected, and that
\be
C_{3} = -12\varepsilon^3 + O(\varepsilon^4)\;,\lb{3.5}
\ee
{\em i.e.} formula (\ref{2.9}) with $n=3$ is false at the lowest nontrivial order in perturbation theory, since it tells that $C_{3} = C_4/2 + O(\varepsilon^5)$, and $C_4 = O(\varepsilon^4)$. This is enough to say that (\ref{2.1}) is violated; in fact, from (\ref{2.3}) it follows that, \cite{tesi},
\be
\z(p) = \frac{(p-1)^{2}}{2}\left[\s_+ - \frac{C_2}{4} + O(\varepsilon^4)\right] - \frac{(p-1)^{3}}{48}\left[C_{3} + O(\varepsilon^4)\right] + O((p-1)^4\varepsilon^4)\;,\lb{3.6}
\ee
that is
\be
-\z(p) + \z(-p) = p\s_+ + p\left[\s_+ - \frac{C_2}{2} + \frac{C_3}{8} + O(\varepsilon^4)\right] + p^3\left[\frac{C_3}{24} + O(\varepsilon^4)\right] + O(p^{5}\varepsilon^5)\;,\lb{3.7}
\ee
and (\ref{3.5}) implies that the difference $-\z(p) + \z(-p)$ is not linear in $p$ (the cubic term is nonzero). This result also shows that, as expected, $I_{\varepsilon}$ {\em cannot be differentiable}: a check that would be not so easy without using the FT.
\section{Conclusions and comparison with \cite{LS}}\lb{sec4}

The Gallavotti --  Cohen Fluctuation Theorem implies nontrivial
identities between correlation functions valid at non zero forcing
$\ul{G}$, which reduce to the usual linear response in the limit $\ul
G\rightarrow \ul 0$; we have shown that it is essential that the time
reversal transformation be (continuously) differentiable. By checking
the FT relation through these identities, in a simple Anosov system
where the time reversal symmetry operator exists but is not
differentiable, we have found that the relation expected from the
possible validity of FT cannot hold (while the linear response is still valid, as expected). Notice, however, that in physical applications the time reversal symmetry operation is usually regular; for instance, it can correspond to the inversion of velocities (or to more subtle permutations of coordinates, see \cite{spontbreak}).

To conclude, it is interesting to note that in presence of a non differentiable time reversal symmetry operator $I$ a different FT holds, the Lebowitz -- Spohn one, see \cite{LS}; strictly speaking this theorem has been proved in the context of general Markov processes, but it can be understood also in the case of deterministic chaotic dynamics. This result applies in particular to systems that are (suitably small) perturbations of reversible Anosov ones, since, as it has been pointed out in the previouos section, by the structural stability of Anosov dynamics the time reversal symmetry is not destroyed by the perturbation (although it will be in general only H\"older continuous in $x$).
 
The fact that a fluctuation relation still holds is a natural consequence of the Gibbian nature of the invariant measure describing the steady state; this has been pointed out in \cite{BGGtest}, where a FT for the one dimensional Ising model in an external field was derived, and then systematized in \cite{Ma}, where large deviation rules for Gibbs states involving transformations different from time reversal were discussed.

Following step by step the proof of Gallavotti -- Cohen FT, \cite{GdimFT}, it is easy to see that the large deviation functional $\tilde \z(\tilde p)$ of the dimensionless quantity $\tilde p = \frac{1}{\t\tilde \s_+}\sum_{j=-\frac{\t}{2}}^{\frac{\t}{2}-1}\tilde\s (S^{j} x)$ where $\tilde \s(x) = -\l_u(x) + \l_u(I x)$ and $\l_u(x)$ is the sum of the positive Lyapunov exponents corresponding to the local unstable manifold $W_u(x)$ of $S$ verifies
\be
\tilde\z(-\tilde p) = \tilde\z(\tilde p) - \tilde p\tilde\s_{+}\;,\lb{4.1}
\ee
(of course under the physical restriction on $\tilde p$ to vary within
the analyticity interval of $\tilde\zeta$, see \cite{GdimFT})
which reduces to (\ref{2.3}) if $I$ is differentiable, since in this case $\l_u(Ix) = -\l_s(x)$. Formula (\ref{4.1}) is equivalent to the Lebowitz -- Spohn fluctuation theorem; in fact, $\tilde\s(x)$ is proportional to the logarithm of the ratio of the SRB probabilities of the trajectories $x_{-\frac{\t}{2}},x_{-\frac{\t}{2}+1},...,x_{\frac{\t}{2}-1}$ and of its time reversed, which is precisely the ``action functional'' introduced in \cite{LS}. 

\bibliographystyle{unsrt}
\bibliography{bibliografia}

\end{document}